\documentclass[showpacs,pra,twocolumn,amsmath,superscriptaddress]{revtex4}
\usepackage{graphicx,amssymb}
\usepackage{bm}
\newcommand{\ket}[1]{\left|#1\right\rangle}
\newcommand{\bra}[1]{\left\langle#1\right|}
\newcommand{\op}[2]{\ket{#1}\bra{#2}}
\newcommand{\ex}[1]{\langle#1\rangle}
\newcommand{\proj}[1]{\op{#1}{#1}}
\newcommand{\com}[2]{\left[#1,#2\right]}
\newcommand{\dampingnobr}[3]{2#1#2#3-#3#1#2-#2#3#1} 
\newcommand{\damping}[3]{\brar{\dampingnobr{#1}{#2}{#3}}}
\newcommand{\modulus}[1]{\left|#1\right|}

\newcommand{\brar}[1]{\left(#1\right)}

\begin{document}

\title{Proposed realization of the Dicke-model quantum phase transition\\ in an optical cavity QED system}
\author{F. Dimer}
\affiliation{Department of Physics, University of Auckland,
Private Bag 92019, Auckland, New Zealand.}
\author{B. Estienne}
\affiliation{Laboratoire de Physique Th\'eorique et Hautes Energies, Universit\'e Pierre et Marie
Curie, 4 place Jussieu, F-75252 Paris Cedex 05, France}
\author{A.~S. Parkins}
\altaffiliation[Permanent address: ]{Department of Physics, University of Auckland, 
Private Bag 92019, Auckland, New Zealand}
\affiliation{Norman Bridge Laboratory of Physics 12-33, California Institute of Technology,
Pasadena, CA 91125, U.S.A.}
\author{H.~J. Carmichael}
\affiliation{Department of Physics, University of Auckland,
Private Bag 92019, Auckland, New Zealand.}
\date{\today}

\begin{abstract}
The Dicke model consisting of an ensemble of two-state atoms interacting with a single  quantized mode of the
electromagnetic field exhibits a zero-temperature phase transition at a critical value of the dipole coupling
strength. We propose a scheme based on multilevel atoms and cavity-mediated Raman transitions to realise an
effective Dicke system operating in the phase transition regime. Output light from the cavity carries signatures
of the critical behavior which is analyzed for the thermodynamic limit where the number of atoms is very large.
\end{abstract}
\pacs{03.65.Ud, 42.50.-p, 42.50.Fx, 05.70.Fh}
\maketitle

\section{Introduction}

The interaction of an ensemble of $N$ two-level atoms with a single mode of the electromagnetic field is a
classic problem in quantum optics and continues to provide a fascinating avenue of research in a variety of
contexts. The simplest model of this interaction is provided by the Dicke Hamiltonian \cite{Dicke54}, which
takes the form ($\hbar=1$)
\begin{equation}
\hat{H}=\omega\hat{a}^\dagger\hat{a}+\omega_0\hat{J}_z+\frac{\lambda}{\sqrt{N}}\left(\hat{a}^\dagger+\hat{a}
\right)\left(\hat{J}_++\hat{J}_-\right),
\label{eq:DickeH}
\end{equation}
where $\omega_0$ is the frequency splitting between the atomic levels, $\omega$ is the frequency of the field
mode, and $\lambda$ is the dipole coupling strength. The boson operators $\{ \hat{a},\hat{a}^\dagger\}$ are
annihilation and creation operators for the field, and $\{\hat{J}_\pm,\hat{J}_z\}$ are collective atomic
operators satisfying angular momentum commutation relations,
\begin{equation}
\left[\hat{J}_+,\hat{J}_-\right]=2\hat{J}_z ,\qquad\left[\hat{J}_\pm,\hat{J}_z\right]=\mp\hat{J}_\pm.
\label{eq:comrels}
\end{equation}
Contained within the possible solutions to this model are a number of significant and topical phenomena, 
including:
\begin{enumerate}
\item[(i)]
A zero-temperature phase transition in the thermodynamic limit, $N\rightarrow\infty$, occurring at the critical
coupling strength $\lambda_{\rm c}=\sqrt{\omega\omega_0}/2$. For larger than the critical coupling the system
enters a super-radiant phase \cite{Hepp73a,Hepp73b,Wang73,Hioe73,Carmichael73,Duncan74,Emary03a,Emary03b}.
\item[(ii)]
An associated change in level statistics, indicating a change from ``quasi-integrable'' to
``quantum chaotic'' behavior \cite{Emary03a,Emary03b}.
\item[(iii)]
Critical behavior of the atom-field entanglement, which diverges at the critical point for $N\rightarrow\infty$
\cite{Lambert04a,Lambert05,Reslen05}.
\end{enumerate}
It follows that the Dicke model offers a potential setting for investigations of quantum critical behavior,
quantum chaos, and quantum entanglement.

Practical realization of a system exhibiting the mentioned phenomena presents something of a problem, however,
in that, in familiar quantum-optical systems, the frequencies $\omega$ and $\omega_0$ typically exceed the dipole
coupling strength by many orders of magnitude. This means that the counter-rotating terms, $\hat{a}^\dagger
\hat{J}_+$ and $\hat{a}\hat{J}_-$ in Eq.~(\ref{eq:DickeH}), have very little effect on the dynamics; indeed,
they are usually neglected in the so-called ``rotating-wave approximation.'' Furthermore, dissipation due to
atomic spontaneous emission and cavity loss is usually unavoidable, significantly altering the pure Hamiltonian
evolution. Hence, it remains as a challenge to provide a practical physical system which might exhibit the
interesting behavior associated with the idealized Dicke model.

We propose such a physical system  in this paper. We suggest a scheme based on interactions in cavity quantum
electrodynamics (cavity QED) which realizes an effective Dicke Hamiltonian (\ref{eq:DickeH}) with parameters 
$\omega_0\simeq\omega\simeq\lambda$ that are adjustable and can in principle far exceed all dissipation rates.
Our scheme uses cavity-plus-laser mediated Raman transitions between a pair of stable atomic ground states,
thereby avoiding spontaneous emission. While cavity loss cannot be similarly avoided, it should be  possible
to achieve cavity QED conditions in which the dissipation rate from the cavity mode is much less than the
parameters of the Dicke model.

In fact, the presence of cavity loss constitutes an important and essential aspect of the work presented here:
output light from the cavity provides a readily measurable signal from which an experimenter can learn, rather
directly, about the properties of the system. In particular, various spectral measurements made on the output
light clearly reveal the critical behavior of the Dicke model as the coupling parameter is changed.

We begin in Sec.~II with a description of the proposed scheme for realizing the Dicke model in an optical
cavity QED system. In Sec.~III, we briefly discuss a possible experimental scenario involving atoms confined
within a ring cavity and establish parameter values for use in the numerical calculations. Our theoretical
study of the dissipative Dicke model in the thermodynamic limit is presented in Sec.~IV. It is based upon
a linearized analysis in the Holstein-Primakoff representation of the collective atomic spin and the input-output
theory of open quantum-optical systems. We present results for the cavity fluorescence spectrum, the probe 
transmission spectrum, and the spectra of quadrature fluctuations---i.e., homodyne, or squeezing spectra.
These spectra vividly illustrate the changing nature of the system through the critical region of the
phase transition. We also describe a means of computing variance-based measures of atom-field entanglement
from homodyne spectra of the cavity output field. We finish in Sec.~V with the conclusion and a discussion
of possible further investigations.

\section{Proposed realization: balanced Raman channels}

\begin{figure}[t]
\begin{center}
\includegraphics[scale=0.35]{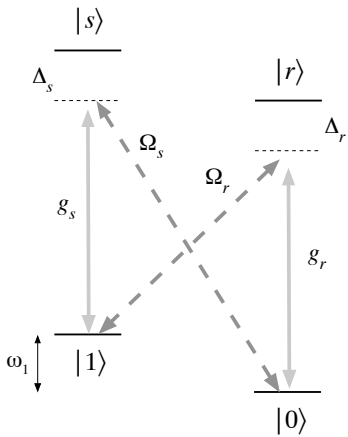}
\caption{
Atomic level scheme. Excited states have energies $\hbar\omega_j$ ($j=r,s$). Such a scheme might be realized,
e.g., by alkali atoms, with $|0\rangle$ and $|1\rangle$ as different ground-state sublevels. Note that $|r\rangle$
and $|s\rangle$ may be the same level, provided that the Raman channels remain distinct (which requires
$\omega_1\neq 0$).}
\label{fig:levels}
\end{center}
\end{figure}

We consider an ensemble of $N$ atoms coupled simultaneously to the quantized field of an optical cavity mode and
the classical field of a pair of lasers. All fields are co-propagating (in the $x$ direction) TEM$_{00}$ traveling
waves, with beam waists sufficiently broad compared to the atomic ensemble that uniform atom-field coupling strengths
may be assumed. Each atom has two stable ground states, $\ket{0}$ and $\ket{1}$, which are coupled through a pair of
Raman channels, as shown in Fig.~\ref{fig:levels}; specifically, the lasers drive ground-to-excited-state transitions
$\ket{1}\leftrightarrow\ket{r}$ and $\ket{0}\leftrightarrow\ket{s}$ with Rabi frequencies $\Omega_r$ and $\Omega_s$,
respectively, while the cavity mode mediates the $\ket{r}\leftrightarrow\ket{0}$ and $\ket{s}
\leftrightarrow\ket{1}$ transitions, with dipole coupling strengths $g_r$ and $g_s$. The detunings from the
excited states are $\Delta_r$ and $\Delta_s$, as shown on the figure.

With the inclusion of spontaneous emission and cavity loss, the master equation for the system density operator,
$\rho_{\rm sys}$, is written as
\begin{equation}
\dot{\rho}_{\rm sys}=-i\com{\hat{H}_{\rm sys}}{\rho_{\rm sys}}+\mathcal{L}_{\rm cav}\rho_{\rm sys}
+\mathcal{L}_{\rm spon}\rho_{\rm sys},
\end{equation}
where $\hat{H}_{\rm sys}$ is a sum of Hamiltonians:
\begin{subequations}
\begin{eqnarray}
\hat{H}_{\rm cav}=\omega_{\rm cav}\hat{a}^\dag \hat{a},
\end{eqnarray}
for the cavity oscillator,
\begin{widetext}
\begin{eqnarray}
\hat{H}_{\rm at}&=&\sum_{j=1}^N \left\{\right.
\omega_r\proj{r_j}+\omega_s\proj{s_j}+\omega_1\proj{1_j}
\nonumber\\
&&+\left.\left[(\Omega_{r}/2)e^{-i\omega_{lr}t}\op{r_j}{1_j} e^{ik_rx_j}+(\Omega_{s}/2) e^{-i\omega_{ls}t}\op{s_j}
{0_j}e^{ik_sx_j}+\textrm{H.c.}\right]\right\},
\end{eqnarray}
for the driven atoms (H.c. denotes the Hermitian conjugate), and
\begin{eqnarray}
\hat{H}_{\rm int}&=&\sum_{j=1}^N \left[\left(g_r\op{r_j}{0_j}\hat{a}+g_s\op{s_j}{1_j}\hat{a}\right)e^{ikx_j}
+\textrm{H.c.}\right],
\end{eqnarray}
\end{widetext}
\end{subequations}
for the atom-cavity interaction, where $\omega_r$, $\omega_s$, and $\omega_1$ are atomic frequencies (see
Fig.~\ref{fig:levels}), $\omega_{lr}$ and $\omega_{ls}$ are the laser frequencies, and $x_j$ locates the $j$-th
atom in the traveling waves, which have wavenumbers $k_r$, $k_s$, and $k$
(where $k_r\simeq k_s\simeq k$).
Cavity loss is included through the Lindblad term
\begin{eqnarray}
\mathcal{L}_{\rm cav}\rho_{\rm sys}&=& \kappa\damping{\hat{a}}{\rho_{\rm sys}}{\hat{a}^\dag},
\label{eq:damping}
\end{eqnarray}
and spontaneous emission through the second Lindblad term $\mathcal{L}_{\rm spon}\rho_{\rm sys}$.

From this full master equation a simplified equation is derived by neglecting spontaneous emission and adiabatically
eliminating the atomic excited states. We first transform to the interaction picture, introducing the unitary
transformation $\hat{U}(t)=\exp(-i\hat{H}_0t)$,
with
\begin{widetext}
\begin{eqnarray}
\hat{H}_0=(\omega_{ls}-\omega_1^\prime)\hat{a}^\dagger \hat{a}+\sum_{j=1}^N\left\{(\omega_{lr}+\omega_1^\prime)
\op{r_j}{r_j}+\omega_{ls}\op{s_j}{s_j}+\omega_1^\prime\op{1_j}{1_j} \right\},
\end{eqnarray}
\end{widetext}
where $\omega_1^\prime$ is a frequency close to $\omega_1$, satisfying
\begin{equation}
\omega_{ls}-\omega_{lr}=2\omega_1^\prime.
\end{equation}
Then assuming large detunings of the fields from the excited states,
\begin{equation}
\modulus{\Delta_{r,s}}\gg\Omega_{r,s},\,g_{r,s},\,\kappa,\,\delta_{\rm cav},\,\gamma,
\end{equation}
where $\gamma$ the excited state linewidth and
\begin{eqnarray}
\Delta_r&=&\omega_r-\left(\omega_{lr}+\omega_1^\prime\right),\qquad\Delta_s=\omega_s-\omega_{ls},\\
&&\quad\delta_{\rm cav}=\omega_{\rm cav}-\left(\omega_{ls}-\omega_1^\prime\right),
\end{eqnarray}
we make the adiabatic elimination and neglect constant energy terms to arrive at the simplified master equation
for the collective coupling of the ground states $\ket{0}$ and $\ket{1}$,
\begin{eqnarray}
\dot{\rho}=-i\left[ \hat{H},\rho\right]+{\cal L}_{\rm cav}\rho,
\label{eq:ME}
\end{eqnarray}
with
\begin{eqnarray}
\hat{H}&=&\omega \hat{a}^\dagger \hat{a}+\omega_0 \hat{J}_z+\delta\hat{a}^\dagger \hat{a}\hat{J}_z+\frac{\lambda_r}
{\sqrt{N}}\left(\hat{a}\hat{J}_++\hat{a}^\dagger \hat{J}_-\right) 
\nonumber\\
&&+\frac{\lambda_s}{\sqrt{N}}\left(\hat{a}^\dagger\hat{J}_++\hat{a}\hat{J}_-\right),
\end{eqnarray}
where
\begin{subequations}
\begin{eqnarray}
&&\hat{J}_+\equiv\sum_{j=1}^N \op{1_j}{0_j},\qquad \hat{J}_-\equiv \sum_{j=1}^N \op{0_j}{1_j},\\
&&\hat{J}_z\equiv\frac{1}{2}\sum_{j=1}^N \left( \op{1_j}{1_j}-\op{0_j}{0_j}\right)
\end{eqnarray}
\end{subequations}
are collective atomic operators satisfying commutation relations (\ref{eq:comrels}), and the remaining parameters of
the model are defined by
\begin{subequations}
\begin{eqnarray}
\omega&=&\tfrac{1}{2}N(g_r^2/\Delta_r+g_s^2/\Delta_s)+\delta_{\rm cav},\\
\omega_0&=&\tfrac{1}{4} \left(\Omega_r^2/\Delta_r-\Omega_s^2/\Delta_s \right)+\left(\omega_1-\omega_1^\prime\right),\\
\delta&=&g_r^2/\Delta_r-g_s^2/\Delta_s,\\
\lambda_r&=&\tfrac{1}{2}\sqrt{N}g_r\Omega_r/\Delta_r,\\
\lambda_s &=&\tfrac{1}{2}\sqrt{N} g_s\Omega_s/\Delta_s.
\end{eqnarray}
\end{subequations}
With these parameters chosen such that
\begin{eqnarray}
g_r^2/\Delta_r=g_s^2/\Delta_s,\qquad g_r\Omega_r/\Delta_r=g_s\Omega_s/\Delta_s,
\end{eqnarray}
$\hat{H}$ is put into the form of the Dicke Hamiltonian (\ref{eq:DickeH}),
\begin{eqnarray}
\hat{H}=\omega \hat{a}^\dagger \hat{a}+\omega_0\hat{J}_z+\frac{\lambda}{\sqrt{N}}\left(\hat{a}+\hat{a}^\dagger\right)
\left(\hat{J}_++\hat{J}_-\right) ,
\label{eq:DickeHeff}
\end{eqnarray}
with
\begin{subequations}
\begin{eqnarray}
\omega&=&Ng_r^2/\Delta_r+\delta_{\rm cav},\\
\omega_0&=&\omega_1-\omega_1^\prime,\\
\lambda&=&\tfrac{1}{2}\sqrt{N}g_r\Omega_r/\Delta_r.
\end{eqnarray}
\end{subequations}
Hence, we arrive at a realization of the Dicke model with parameters that can be controlled through the laser frequencies
and intensities, and where the characteristic energy scales are no longer those of optical photons and dipole coupling
but those associated with light shifts and Raman transition rates.

\section{Potential experimental scheme}

\begin{figure}[h]
\begin{center}
\includegraphics[scale=0.3]{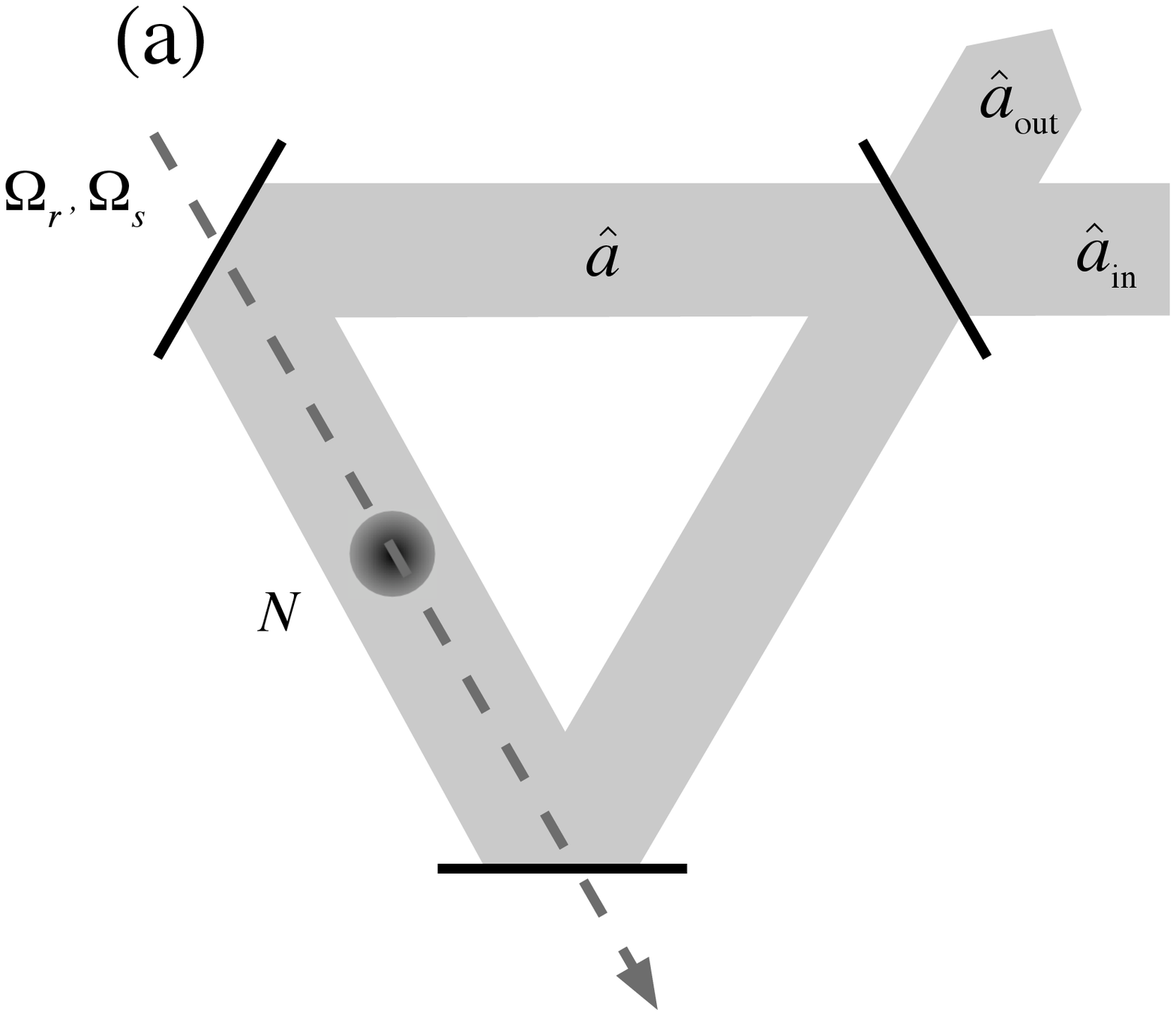}\\
\includegraphics[scale=0.3]{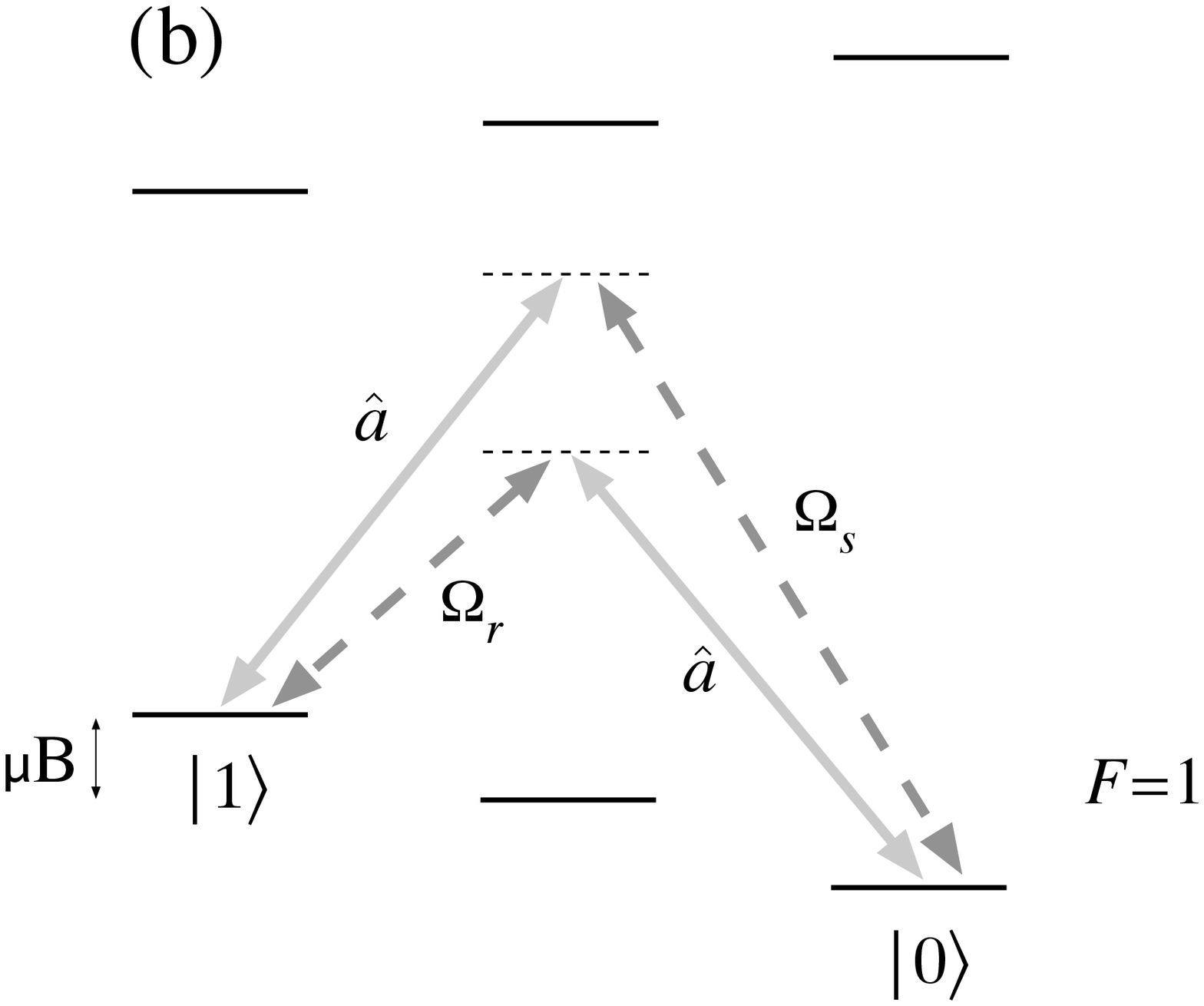}\\
\caption{(a) Ring cavity configuration for implementing the proposed realization of the Dicke model. Quantized input and
output fields are denoted by $\hat{a}_{\rm in}$ and $\hat{a}_{\rm out}$, respectively. (b) Possible atomic excitation
scheme based upon an $F=1\leftrightarrow F^\prime=1$ atomic transition and a linearly-polarized cavity field $\hat{a}$.
Note that the magnetic field splittings of the Zeeman sublevels are not drawn to scale; the detunings of the optical fields from the excited
atomic states are much larger than the ground-state splittings (i.e., $|\Delta_{r,s}|\gg |\mu B|$).}
\label{fig:ringcavity}
\end{center}
\end{figure}

Before proceeding with our theoretical analysis, we pause briefly to outline a possible experimental implementation of the
proposed model. We imagine the ensemble of atoms confined inside a ring cavity where it interacts with the quantized cavity
mode (field operator $\hat a$) as shown in Fig.~\ref{fig:ringcavity}(a). The cavity mode copropagates with the two laser
fields (Rabi frequencies $\Omega_r$ and $\Omega_s$) through the ensemble as indicated on the figure by the dashed line.
Quantized inputs and outputs are assumed significant through one cavity mirror only---field operators $\hat a_{\rm in}$
and $\hat a_{\rm out}$ in the figure.

The atomic excitation scheme might be based on an $F=1\leftrightarrow F^\prime=1$ transition, as occurs, for example,
in ${}^{87}$Rb. Such a scheme differs slightly from that of Fig.~(\ref{fig:levels}) and is illustrated in
Fig.~\ref{fig:ringcavity}(b). The cavity mode is linearly polarized along an axis perpendicular to an applied magnetic
field of strength $B$. The magnetic field splits the $m_F=\pm1$ sublevels of the $F=1$ ground state, allowing for the
excitation of the distinct Raman channels shown \cite{Wilk06}.

Parameter values $g_{r}/2\pi\simeq 50~\textrm{kHz}$, $\kappa/2\pi\simeq 20~\textrm{kHz}$, and $N\simeq10^6$ appear to be
practical \cite{Kruse03,Nagorny03}; thus, with the choice $\Omega_r/\Delta_r=0.005$, one finds an effective coupling
strength $\lambda/2\pi=\frac{1}{2}\sqrt{N}g_r\Omega_r/2\pi \Delta_r\simeq 125~\textrm{kHz}$. This is significantly larger
than the decay rate $\kappa$, placing the system firmly in a regime where the Hamiltonian dynamics can be expected to
dominate. Note further that, for these parameters, the spontaneous emission rate due to off-resonant excitation of the
atomic excited states is estimated at $\frac{1}{4}(\gamma/2\pi)(\Omega_r/\Delta_r)^2\lesssim 40~\textrm{Hz}$, where 
$\gamma/2\pi=6~\textrm{MHz}$ has been assumed. Finally, the condition $\omega\simeq\omega_0\simeq\lambda$ can be achieved
with appropriate choices of the laser and cavity mode frequencies, and ground-state level shifts of the order of
$2\pi\cdot$10-15~MHz ($\gtrsim 100\lambda$) would satisfy the requirement for distinct Raman channels.

The above set of parameters provides just one example of the possibilities, and a wide variety of parameter combinations
satisfy the requirements of our model. In what follows we concentrate in large part, for numerical investigations, 
on the set of (normalized) parameters $\{\omega,\omega_0,\kappa\}=\{ 1,1,0.2\}$. This choice serves to highlight
the main physical features of the model proposed.

\section{Analysis in the Thermodynamic limit}

We aim to make a theoretical analysis of the Dicke-model quantum phase transition in the thermodynamic limit, i.e., for
$N\gg1$. Our starting point is a semiclassical analysis of the steady state and its bifurcations, to which a linearized
treatment of quantum fluctuations is added using the Holstein-Primakov representation and the input-output
theory of open quantum systems.

\subsection{Semiclassical steady states}

\begin{figure}[h]
\begin{center}
\hbox{\hskip-0.25cm\includegraphics[scale=0.5]{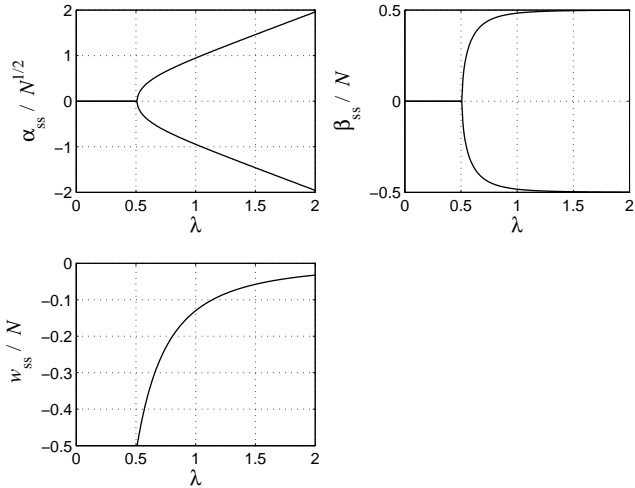}}
\vskip-0.3cm
\caption{Steady-state field amplitude, $\alpha_{\rm ss}$, polarization amplitude, $\beta_{\rm ss}$, and atomic inversion,
$w_{\rm ss}$, plotted as a function of the coupling strength $\lambda$, for $\omega=\omega_0=1$ and $\kappa=0.2$. Only
stable steady states are shown.}
\label{fig:amplitudes}
\end{center}
\end{figure}

Introducing the $c$-number variables
\begin{equation}
\alpha\equiv\langle\hat{a}\rangle,\qquad\beta\equiv\langle\hat{J}_-\rangle,\qquad w\equiv\langle\hat{J}_z\rangle, 
\end{equation}
where $\alpha$ and $\beta$ are the complex field and atomic polarization amplitudes, respectively, and $w$ is the (real)
population inversion, we examine the semiclassical equations of motion
\begin{subequations}
\begin{eqnarray}
\dot{\alpha}&=&-(\kappa +i\omega)\alpha-i\frac{\lambda}{\sqrt{N}}(\beta+\beta^\ast ),
\label{eq:OBEs(a)}\\
\dot{\beta}&=&-i\omega_0\beta+2i\frac{\lambda}{\sqrt{N}}(\alpha +\alpha^\ast)w,
\label{eq:OBEs(s)}\\
\dot{w}&=&i\frac{\lambda}{\sqrt{N}}(\alpha +\alpha^\ast)(\beta -\beta^\ast).
\label{eq:OBEs(c)}
\end{eqnarray}
These follow from master equation (\ref{eq:ME}), with Hamiltonian (\ref{eq:DickeHeff}) and cavity damping (\ref{eq:damping}),
by neglecting quantum fluctuations and imposing the factorization
\end{subequations}
\begin{eqnarray}
\langle\left( \hat{a}+\hat{a}^\dagger \right)\hat{J}_z\rangle&\rightarrow&\langle\left( \hat{a}+\hat{a}^\dagger\right)
\rangle \langle\hat{J}_z\rangle,\nonumber\\
\langle\left(\hat{a}+\hat{a}^\dagger\right)(\hat{J}_--\hat{J}_+)\rangle &\rightarrow&\langle\left(\hat{a}+\hat{a}^\dagger
\right)\rangle\langle (\hat{J}_--\hat{J}_+) \rangle.\nonumber
\end{eqnarray}
The semiclassical equations conserve the magnitude of pseudo angular momentum,
\begin{equation}
w^2+\left|\beta\right|^2=N^2/4.
\end{equation}
We use this conservation law and solve Eqs.~(\ref{eq:OBEs(a)})--(\ref{eq:OBEs(c)}) for the steady states, whence a critical
value of the coupling strength occurs at
\begin{equation}
\lambda=\lambda_{\rm c}\equiv\frac{1}{2}\sqrt{(\omega_0/\omega )(\kappa^2+\omega^2)}.
\end{equation}
For $\lambda <\lambda_{\rm c}$, there are two steady states,
\begin{equation}
\alpha_{\rm ss}=\beta_{\rm ss}=0,\qquad w_{\rm ss}=\pm N/2,
\end{equation}
where the states with negative  and positive inversion are dynamically stable and unstable, respectively. Both states become
unstable for $\lambda>\lambda_{\rm c}$, where the new stable steady states are
\begin{subequations}
\begin{eqnarray}
\alpha_{\rm ss}&=&\pm \sqrt{N}\frac{\lambda}{\omega-i\kappa}\sqrt{1-\lambda_{\rm c}^4/\lambda^4}\, ,
\label{eq:alpha_ss}\\
\beta_{\rm ss}&=&\mp\frac{N}{2} \sqrt{1-\lambda_{\rm c}^4/\lambda^4}\,,
\label{eq:beta_ss}\\
w_{\rm ss}&=&-\frac{N}{2}\lambda_{\rm c}^2/\lambda^2.
\label{eq:w_ss}
\end{eqnarray}
\end{subequations}
These quantities are plotted as a function of the coupling strength in Fig.~\ref{fig:amplitudes}. Note the bifurcation
to states of finite amplitude and inversion at $\lambda=\lambda_c$. This is the Dicke-model quantum phase transition
\cite{Hepp73a,Hepp73b,Wang73,Hioe73,Carmichael73,Duncan74,Emary03a,Emary03b} as encountered, without fluctuations, in the
thermodynamic limit.

\subsection{Linearized treament of quantum fluctuations in the Holstein-Primakoff representation}

In the thermodynamic limit, $N\gg1$, the quantum fluctuations are small and may be treated in a linearized approach. We follow
Emary and Brandes \cite{Emary03a,Emary03b,Lambert04a,Lambert05} and make use of the Holstein-Primakoff representation of
angular momentum operators \cite{Holstein40,Ressayre75}. Collective atomic operators $\hat J_+$, $\hat J_-$, and $\hat J_z$
are expressed in terms of annihilation and creation operators, $\hat{b}$ and $\hat{b}^\dagger$, of a single bosonic mode:
\begin{subequations}
\begin{eqnarray}
\hat{J}_+&=&\hat{b}^\dagger \sqrt{N-\hat{b}^\dagger\hat{b}},\qquad\hat{J}_-=\hat J_+^\dagger,
\label{eq:Hol_Prima}\\
\hat{J}_z&=&\hat{b}^\dagger\hat{b}-N/2,
\label{eq:Hol_Primb}
\end{eqnarray}
\end{subequations}
from which, using $[\hat{b},\hat{b}^\dagger]=1$, the angular momentum commutation relations (\ref{eq:comrels}) are recovered. 
Substituting these expressions into the Dicke Hamiltonian, we expand the resulting expression under the assumption $N\gg1$.
The goal is to achieve a linearization about the semiclassical amplitudes derived above; one must therefore distinguish
between the so-called ``normal'' ($\lambda<\lambda_{\rm c}$) and ``superradiant'' ($\lambda>\lambda_{\rm c}$) phases before
the expansion is made.

\subsubsection{Normal phase ($\lambda<\lambda_{\rm c}$)}

The semiclassical amplitudes $\alpha_{\rm ss}$ and $\beta_{\rm ss}$ are zero and the expansion is made directly upon the
operators as they appear in Eqs.~(\ref{eq:Hol_Prima}) and (\ref{eq:Hol_Primb}). This yields the master equation
\begin{eqnarray}
\dot{\rho}=-i\left[\hat{H}^{(1)},\rho\right]+{\cal L}_{\rm cav}\rho ,
\end{eqnarray}
with the Hamiltonian governing fluctuations (omitting constant terms)
\begin{eqnarray}
\hat{H}^{(1)}=\omega \hat{a}^\dagger\hat{a}+\omega_0 \hat{b}^\dagger\hat{b}+\lambda(\hat{a}^\dagger+\hat{a})
(\hat{b}^\dagger +\hat{b}).
\label{eq:H1}
\end{eqnarray}

\subsubsection{Superradiant phase ($\lambda>\lambda_{\rm c}$)}

The semiclassical amplitudes $\alpha_{\rm ss}$ and $\beta_{\rm ss}$ are nonzero and the expansion of the Hamiltonian is
preceded by making coherent displacements of $\hat a$ and $\hat b$, as both bosonic modes are macroscopically excited.
Specifically, defining
\begin{equation}
\tilde{\mu}=\lambda_{\rm c}^2/\lambda^2<1,
\end{equation}
we make transformations
\begin{equation}
\hat{a}\rightarrow\hat{c}+\alpha_{\rm ss},\qquad\hat{b}\rightarrow\hat{d}+\frac{\beta_{\rm ss}}{\sqrt{(N/2)
(1+\tilde{\mu})}},
\end{equation}
where $\alpha_{\rm ss}$ and $\beta_{\rm ss}$ are given in Eqs.~(\ref{eq:alpha_ss}) and (\ref{eq:beta_ss}), and $\hat c$
and $\hat d$ describe quantum fluctuations about the semiclassical steady state. We then proceed with the expansion to
obtain the master equation
\begin{eqnarray}
\dot{\rho}=-i\left[\hat{H}^{(2)},\rho\right]+{\cal L}_{\rm cav}^\prime\rho,
\end{eqnarray}
with the Hamiltonian governing fluctuations (omitting constant terms)
\begin{widetext}
\begin{eqnarray}
\hat{H}^{(2)}=\omega \hat{c}^\dagger\hat{c}+\frac{\omega_0}{2\tilde{\mu}}\left(1+\tilde{\mu} \right)\hat{d}^\dagger
\hat{d}+\frac{\omega_0(1-\tilde{\mu})(3+\tilde{\mu})}{8\tilde{\mu}(1+\tilde{\mu})}(\hat{d}+\hat{d}^\dagger )^2
+\lambda \tilde{\mu}\sqrt{\frac{2}{1+\tilde{\mu}}}\mkern2mu(\hat{c}^\dagger+\hat{c})(\hat{d}^\dagger+\hat{d}),
\label{eq:H2}
\end{eqnarray}
\end{widetext}
and
\begin{equation}
{\cal L}_{\rm cav}^\prime\rho=\kappa\damping{\hat{c}}{\rho}{\hat{c}^\dag} .
\end{equation}

\subsubsection{Eigenvalue analysis}
\label{sec:eigenvalues}

\begin{figure}[b]
\begin{center}
\hbox{\hskip-0.25cm\includegraphics[scale=0.5]{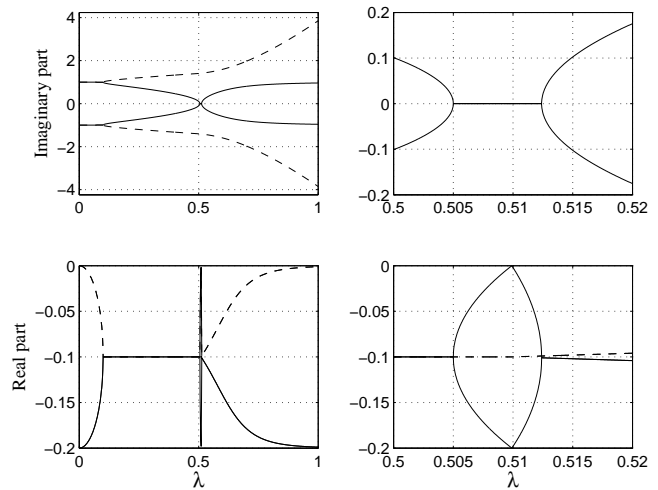}}
\caption{Imaginary parts (upper row) and real parts (lower row) of the eigenvalues in the linearized Holstein-Primakoff
representation as a function of the coupling strength $\lambda$; for $\omega=\omega_0=1$ and $\kappa=0.2$. Solid
(dashed) lines  correspond to the photonic (atomic) branch. The right-hand column magnifies the view around the transition
at $\lambda=\lambda_c=0.5099$; note the splitting (convergence) at $\lambda^\prime\simeq0.5050$ and $\lambda^{\prime\prime}
\simeq 0.5124$.}
\label{fig:eigenvalues}
\end{center}
\end{figure}

The quadratic Hamiltonians and dissipative Lindblad terms above lead to linear equations of motion for the expectation
values of $\hat c$ and $\hat d$. We write
\begin{equation}
\dot{\bm v}={\bm M}{\bm v},
\end{equation}
where ${\bm M}$ is a constant matrix and
\begin{equation}
{\bm v}\equiv\left(\ex{\hat{c}},\ex{\hat{c}^\dagger},\ex{\hat{d}},\ex{\hat{d}^\dagger}\right)^T.
\end{equation}
The eigenvalues of ${\bm M}$ are plotted as a function of coupling strength in Fig.~\ref{fig:eigenvalues} for
$\omega=\omega_0=1$ and $\kappa=0.2$, with the four eigenvalues grouped into pairs, one pair associated with the
``photonic'' branch and the other with the ``atomic''  branch; the branches are defined by the $\lambda\rightarrow 0$
limit of the corresponding eigenstates (or, in fact, the $\lambda\rightarrow\infty$ limit) \cite{Emary03a,Emary03b}.
Note that with the nonzero cavity damping, there are two coupling strengths of significance in addition to
$\lambda_{\rm c}$; for $\omega=\omega_0$ they are
\begin{equation}
\lambda^\prime\simeq\lambda_{\rm c}-\kappa^2/8\omega_0^2,\qquad\lambda^{\prime\prime}\simeq\lambda_{\rm c}
+\kappa^2/16\omega_0,
\end{equation}
with $\lambda^\prime<\lambda_{\rm c}<\lambda^{\prime\prime}$. As $\lambda\rightarrow[\lambda^\prime]_-$ and 
$\lambda\rightarrow[\lambda^{\prime\prime}]_+$ the imaginary parts of the eigenvalues on the photonic branch go to
zero---respectively, as $\sqrt{\lambda^\prime-\lambda}$ from below and $\sqrt{\lambda-\lambda^{\prime\prime}}$ from
above. They remain zero in the interval $\lambda^\prime<\lambda<\lambda^{\prime\prime}$. In correspondence, the real
parts of the eigenvalues split, with the real part of one eigenvalue going to zero at the critical coupling strength
$\lambda_{\rm c}$.

To complement the figure, in the range $0\leq\lambda\leq\kappa/2$, the eigenvalues are given by (with $\omega=\omega_0$)
\begin{subequations}
\begin{eqnarray}
\epsilon_{\rm ph}^\pm&=&-\kappa/2\pm i\sqrt{\omega_0^2-\kappa^2/4\pm\Lambda},\\
\epsilon_{\rm at}^\pm&=&-\kappa/2\pm i\sqrt{\omega_0^2-\kappa^2/4\mp\Lambda}, 
\end{eqnarray}
\end{subequations}
with
\begin{equation}
\Lambda=\sqrt{\omega_0^2(4\lambda^2-\kappa^2)}, 
\end{equation}
where both upper or lower signs are to be taken, while for $\kappa/2\leq\lambda\leq\lambda_{\rm c}$,
\begin{subequations}
\begin{eqnarray}
\epsilon_{\rm ph}^\pm &=&-\kappa/2\pm i\sqrt{\omega_0^2-\kappa^2/4-\Lambda},\\
\epsilon_{\rm at}^\pm &=&-\kappa/2 \pm i\sqrt{\omega_0^2-\kappa^2/4+\Lambda}. 
\end{eqnarray}
\end{subequations}
Thus we see that $\epsilon_{\rm ph}^-\rightarrow 0$, $\epsilon_{\rm ph}^+\rightarrow -\kappa$, and
$\epsilon_{\rm at}^\pm\rightarrow-\kappa/2\pm i\sqrt{2\omega_0^2-\kappa^2/4}$  as the critical coupling is approached. 

Above the critical point, similarly simple expressions cannot be found. We note, however, that for
$\lambda>\lambda^{\prime\prime}$ the photonic branch eigenvalues take on nonzero imaginary parts once again, and for
large $\lambda$ approach $-\kappa\pm i\omega_0$. The atomic branch eigenvalues approach $\pm i\omega_0/\tilde\mu$,
with a rapidly decreasing real part that scales like $\tilde\mu^4=(\lambda_{\rm c}/\lambda)^8$.

\subsection{Quantum Langevin equations and input-output theory}

The equations of motion of the previous sections concern the ``internal'' dynamics of the atom-cavity system. To probe this
dynamics we consider measurements on the light leaving the system through the cavity output mirror. We make use of the
standard input-output theory of open quantum-optical systems \cite{Collett84,Gardiner85,Walls94,Carmichael99}, which is
nicely formulated in terms of quantum Langevin equations for system operators: for $\lambda<\lambda_{\rm c}$,
\begin{subequations}
\begin{eqnarray}
\dot{\hat{a}}&=&-i\left[ \hat{a},\hat{H}^{(1)}\right]
-\kappa\hat{a} + \sqrt{2\kappa}\,\hat{a}_{\rm in}(t),\\
\dot{\hat{b}} &=& -i\left[ \hat{b},\hat{H}^{(1)}\right] ,
\end{eqnarray}
\end{subequations}
plus the adjoint equations, while for $\lambda>\lambda_{\rm c}$,
\begin{subequations}
\begin{eqnarray}
\dot{\hat{c}}&=&-i\left[\hat{c},\hat{H}^{(2)}\right]
-\kappa\hat{c} + \sqrt{2\kappa}\,\hat{a}_{\rm in}(t),\\
\dot{\hat{d}}&=&-i\left[\hat{d},\hat{H}^{(2)}\right],
\end{eqnarray}
\end{subequations}
plus the adjoint equations. The operator $\hat{a}_{\rm in}(t)$ describes the quantum noise injected at the cavity input
(Fig.~\ref{fig:ringcavity}) and satisfies the commutation relation
\begin{equation}
\left[\hat{a}_{\rm in}(t),\hat{a}_{\rm in}^\dagger(t^\prime)\right]=\delta (t-t^\prime).
\end{equation}
In addition, for  vacuum or coherent state inputs, one has the correlations
\begin{subequations}
\begin{eqnarray}
\langle\hat{a}_{\rm in}(t),\hat{a}_{\rm in}^\dagger(t^\prime)\rangle&=&\delta(t-t^\prime),
\label{eq:a_in_corra}\\
\langle\hat{a}_{\rm in}^\dagger(t),\hat{a}_{\rm in}(t^\prime)\rangle&=&\langle\hat{a}_{\rm in}(t),\hat{a}_{\rm in}
(t^\prime)\rangle=0,
\label{eq:a_in_corrb}
\end{eqnarray}
\end{subequations}
where
$\langle\hat{A},\hat{B}\rangle\equiv\langle\hat{A}\hat{B}\rangle-\langle\hat{A}\rangle\langle\hat{B}\rangle$. The cavity
output field, $\hat{a}_{\rm out}(t) $, is given in terms of the intracavity and cavity input fields as
\begin{equation}
\hat{a}_{\rm out}(t)=\sqrt{2\kappa}\,\hat{a}(t)-\hat{a}_{\rm in}(t),
\label{eq:io_relation}
\end{equation}
from which one calculates the output field correlation functions and spectra.

The quantum Langevin equations are linear operator equations. For the purpose of computing spectra, they are conveniently
solved in frequency space by introducing the Fourier transforms 
\begin{subequations}
\begin{eqnarray}
\tilde{\cal O}(\nu)&=&\frac{1}{\sqrt{2\pi}}\int_{-\infty}^\infty e^{i\nu t}\hat{\cal O}(t)\,{\rm d}t,\\
\tilde{\cal O}^\dagger(-\nu)&=&\frac{1}{\sqrt{2\pi}}\int_{-\infty}^\infty e^{i\nu t}\hat{\cal O}^\dagger (t)\,
{\rm d}t,
\end{eqnarray}
\end{subequations}
where $\hat{\cal O}$ denotes any one of the operators $\hat a$, $\hat b$, $\hat c$, $\hat d$, or $\hat a_{\rm in}$. 
In the resonant case, $\omega_0=\omega$, the solutions are: for $\lambda<\lambda_{\rm c}$, 
\begin{widetext}
\begin{subequations}
\begin{eqnarray}
\tilde{a}(\nu) &=& \sqrt{2\kappa}\, \frac{\left\{\left[\kappa -i(\nu +\omega_0)\right]\left(\nu^2-\omega_0^2
\right)-2i\omega_0\lambda^2\right\}\tilde{a}_{\rm in}(\nu)-2i\omega_0\lambda^2\tilde{a}_{\rm in}^\dagger(-\nu)}
{\left[\kappa -i(\nu -\omega_0)\right]\left[\kappa -i(\nu +\omega_0)\right]\left(\nu^2-\omega_0^2\right)+4\omega_0^2
\lambda^2},
\label{eq:a_nu}\\
\tilde{b}(\nu)&=&\frac{\lambda}{\nu -\omega_0}\left[\tilde{a}(\nu)+\tilde{a}^\dagger(-\nu)\right],
\end{eqnarray}
\end{subequations}
and for $\lambda>\lambda_{\rm c}$,
\begin{subequations}
\begin{eqnarray}
\tilde{c}(\nu)&=&\sqrt{2\kappa}\,\frac{\left\{\left[\kappa-i(\nu+\omega_0)\right]\left(\nu^2-\omega_0^2/
\tilde{\mu}^2\right)-2i\omega_0\lambda^2\tilde{\mu}\right\}\tilde{a}_{\rm in}(\nu) -2i\omega_0\lambda^2\tilde{\mu}\,
\tilde{a}_{\rm in}^\dagger(-\nu)}{\left[\kappa-i(\nu-\omega_0)\right]\left[\kappa-i(\nu +\omega_0)\right]
\left(\nu^2-\omega_0^2/\tilde{\mu}^2\right)+4\omega_0^2\lambda^2\tilde{\mu}},
\label{eq:c_nu}\\
\tilde{d}(\nu)&=&\frac{\lambda\tilde{\mu}\sqrt{2/(1+\tilde{\mu})}}{\nu-\omega_0(1+\tilde{\mu})/(2\tilde{\mu})} 
\left[\tilde{c}(\nu)+\tilde{c}^\dagger(-\nu)\right].
\end{eqnarray}
\end{subequations}
\end{widetext}

\subsection{Entanglement of the atoms and field}

\begin{figure}[t]
\begin{center}
\includegraphics[scale=0.4]{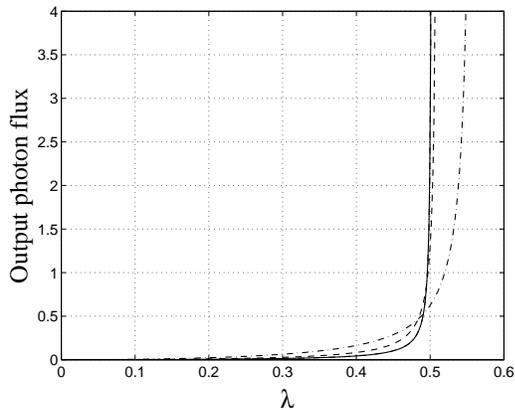}\\
\caption{Output photon flux as a function of coupling strength; for $\omega=\omega_0=1$ and $\kappa=0.1$ (solid),
$0.2$ (dashed), and $0.5$ (dot-dashed).}
\label{fig:photonnumber}
\end{center}
\end{figure}

Quantum fluctuations in the linearized treatment are Gaussian, and the solutions to the quantum Langevin equations can
be used to compute their covariances in the steady state. For example, the mean intracavity photon number for
$\lambda<\lambda_{\rm c}$ is given by
\begin{equation}
\langle\hat{a}^\dagger\hat{a}\rangle_{\rm ss}=\frac{1}{2\pi}\int_{-\infty}^\infty\int_{-\infty}^\infty
\langle\hat{a}^\dagger(\nu)\hat{a}(\nu^\prime)\rangle\,{\rm d}\nu\,{\rm d}\nu^\prime,
\end{equation}
where we need the frequency-space equivalents of the input correlations (\ref{eq:a_in_corra}) and (\ref{eq:a_in_corrb}),
i.e.,
\begin{subequations}
\begin{eqnarray}
\langle\tilde{a}_{\rm in}(\nu),\tilde{a}_{\rm in}^\dagger(\nu^\prime)\rangle&=&\delta(\nu-\nu^\prime),
\label{eq:cor_nua}\\
\langle\tilde{a}_{\rm in}^\dagger(\nu),\tilde{a}_{\rm in}(\nu^\prime)\rangle&=&\langle\tilde{a}_{\rm in}(\nu),
\tilde{a}_{\rm in}(\nu^\prime)\rangle=0.
\label{eq:cor_nub}
\end{eqnarray}
\end{subequations}
The computed output photon flux from the cavity, $2\kappa\langle\hat{a}^\dagger\hat{a}\rangle_{\rm ss}$, is plotted
 for several different values of $\kappa$ in Fig.~\ref{fig:photonnumber}, illustrating a ``smoothing-out'' of the
transition with increasing cavity linewidth. The mean excitation of the atomic mode, $\langle\hat{b}^\dagger
\hat{b}\rangle_{\rm ss}$, shows similar behavior.

Of particular interest is the behavior of the bipartite quantum entanglement in the vicinity of the critical point
\cite{Lambert04a,Lambert05,Reslen05,Osterloh02,Osborne02,Vidal03,Vidal04,Wu04,Latorre05}. The cavity and atomic
modes are natural choices for the entangled subsystems, and given that their fluctuations are described by a Gaussian
continuous variable state, the criterion for inseparability can be formulated in terms of the variances of appropriate
subsystem operators. In particular, we define the quadrature operators
\begin{subequations}
\begin{eqnarray}
\hat{X}_a^\theta&=&\frac{1}{2}\left(\hat{a}e^{-i\theta}+\hat{a}^\dagger e^{i\theta}\right),\\
\hat{X}_b^\phi&=&\frac{1}{2}\left(\hat{b}e^{-i\phi}+\hat{b}^\dagger e^{i\phi}\right),
\end{eqnarray}
\end{subequations}
with adjustable phases $\theta$ and $\phi$, and introduce the EPR (Einstein-Podolsky-Rosen) operators 
\begin{eqnarray}
\hat{u}=\hat{X}_a^\theta+\hat{X}_b^\phi,\qquad\hat{v}=\hat{X}_a^{\theta+\pi/2}-\hat{X}_b^{\phi+\pi/2}.
\end{eqnarray}
Then a sufficient condition for the inseparability of the state below the critical point is (any $\theta$ and $\phi$)
\cite{Duan00}
\begin{equation}
\langle(\Delta\hat{u})^2\mkern1mu\rangle+\langle(\Delta\hat{v})^2\mkern1mu\rangle<1 .
\label{eq:insep}
\end{equation}
Alternatively, a stronger condition may be given in the modified form 
\cite{Giovannetti03}
\begin{equation}
\langle(\Delta\hat{u})^2\mkern1mu\rangle\langle(\Delta\hat{v})^2\mkern1mu\rangle<\frac{1}{4}.
\label{eq:insep2}
\end{equation}
Above the critical point, similar definitions and conditions based on operators $\hat c$ and $\hat d$ hold.
Here the EPR variance is associated with a quantum state ``localized'' about one of the two possible semiclassical
steady states (\ref{eq:alpha_ss})--(\ref{eq:w_ss}); within our linearized treatment transitions between these states
are ignored.

\begin{figure}[t]
\begin{center}
\includegraphics[scale=0.4]{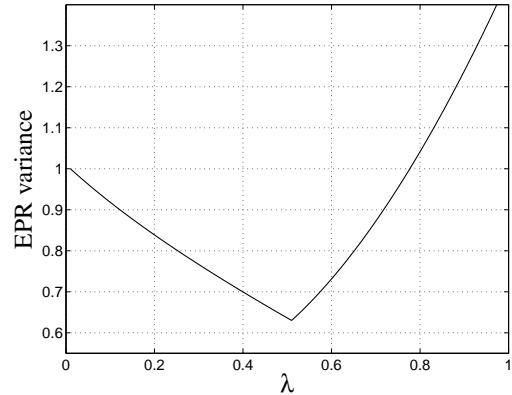}\\
\caption{Sum of EPR operator variances as a function of coupling strength; for $\omega=\omega_0=1$, $\kappa=0.2$,
$\theta=\tan^{-1}(\kappa/\omega)$, and $\phi=0$. The choice of $\theta$ minimizes the sum in the vicinity of the
critical coupling strength, $\lambda=\lambda_{\rm c}\simeq 0.51$.}
\label{fig:EPRvariance}
\end{center}
\end{figure}

The sum of EPR operator variances---inequality (\ref{eq:insep})---is plotted as a function of the coupling strength
in Fig.~\ref{fig:EPRvariance}. It approaches a cusp-like minimum at the critical coupling strength; thus, the
entanglement here is maximum. The variance product---inequality (\ref{eq:insep2})---exhibits similar behavior.
These variances are measurable quantities. They offer a means of tracking entanglement across the phase transition.
In fact, as we show in Sec.~\ref{sec:squeeze}, variance-based entanglement measures can, under appropriate conditions,
be deduced from measurements on the cavity output field alone.

\subsection{Spectra of the cavity output field}

Cavity output field spectra can be computed from the solutions to the Langevin equations (\ref{eq:a_nu}) and
(\ref{eq:c_nu}), the correlations (\ref{eq:cor_nua}) and (\ref{eq:cor_nub}), and the
input-output relations
\begin{equation}
\tilde{a}_{\rm out}(\nu)=\sqrt{2\kappa}\,\tilde{a}(\nu)-\tilde{a}_{\rm in}(\nu),
\end{equation}
$\lambda<\lambda_{\rm c}$, and
\begin{equation}
\tilde{a}_{\rm out}(\nu)=\sqrt{2\kappa}\,\left[ \tilde{c}(\nu)+\sqrt{2\pi}\,\alpha_{\rm ss}\delta(\nu)\right]
-\tilde{a}_{\rm in}(\nu),
\end{equation}
$\lambda>\lambda_{\rm c}$. We consider three standard spectra: (i) the fluorescence (or power) spectrum, which
is proportional to the probability of detecting a photon of frequency $\nu$ at the cavity output, (ii) the probe
transmission spectrum, the transmitted intensity as a function of frequency of a (weak) probe field applied
at the cavity input, and (iii) homodyne spectra, which measure the quantum noise variances of output field
quadrature amplitudes.

\subsubsection{Fluorescence spectrum}
The fluorescence spectrum consists of a coherent part, representing the mean excitation of the intracavity field,
the semiclassical solution $\alpha_{\rm ss}$, and an incoherent part which accounts for the quantum fluctuations.
The latter is defined by
\begin{eqnarray}
\langle\tilde{a}_{\rm out}^\dagger(\nu),\tilde{a}_{\rm out}(\nu^\prime)\rangle=S(\nu)\delta(\nu-\nu^\prime).
\end{eqnarray}
It can also be expressed in terms of the steady state autocorrelation function of the intracavity field, with
\begin{eqnarray}
S^{(1)}(\nu)=\int_{-\infty}^\infty e^{-i\nu\tau}\langle\hat{a}^\dagger(\tau),\hat{a}(0)\rangle_{\rm ss}
{\rm d}\tau,
\end{eqnarray}
$\lambda<\lambda_{\rm c}$, and
\begin{eqnarray}
S^{(2)}(\nu)=\int_{-\infty}^\infty e^{-i\nu\tau}\langle\hat{c}^\dagger(\tau),\hat{c}(0)\rangle_{\rm ss}
{\rm d}\tau,
\end{eqnarray}
$\lambda>\lambda_{\rm c}$.
Making use of solutions (\ref{eq:a_nu}) and (\ref{eq:c_nu}) for $\tilde{a}(\nu)$ and $\tilde{c}(\nu)$, and
the input correlations (\ref{eq:cor_nua}) and (\ref{eq:cor_nub}), one finds
\begin{widetext}
\begin{eqnarray}
S^{(1,2)}(\nu)=\left|\frac{4\kappa\omega_0\lambda^2\tilde{\mu}^{(1,2)}}{\left[\kappa -i(\nu-\omega_0)\right]
\left[\kappa-i(\nu+\omega_0)\right]\left[\nu^2-\omega_0^2/\left(\tilde{\mu}^{(1,2)}\right)^2\right]
+4\omega_0^2\lambda^2\tilde{\mu}^{(1,2)}}\right|^2 , 
\end{eqnarray}
\end{widetext}
with the definitions
\begin{eqnarray}
\tilde{\mu}^{(1)}=1,\qquad \tilde{\mu}^{(2)}=\lambda_{\rm c}^2/\lambda^2.
\label{eq:mu12}
\end{eqnarray}

\begin{figure}[h]
\begin{center}
\includegraphics[scale=0.4]{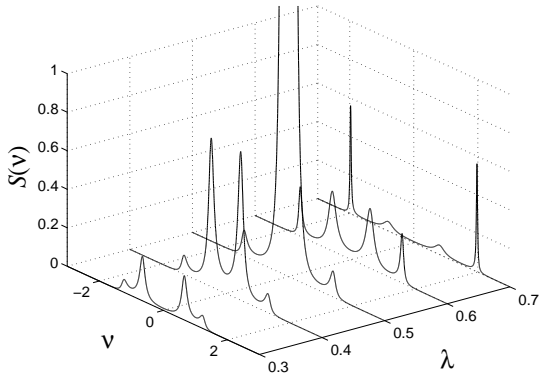}
\caption{Incoherent part of the cavity fluorescence spectrum $S(\nu)$ for various values of coupling strength
$\lambda$; for $\omega=\omega_0=1$ and $\kappa=0.2$ ($\lambda_{\rm c}\simeq0.51$).}
\label{fig:fluorescence}
\end{center}
\end{figure}

Sample spectra $S^{(1,2)}(\nu)$ are plotted in Fig.~\ref{fig:fluorescence}. The positions and widths of the
spectral peaks are determined by the eigenvalues of the linearized dynamics discussed in Sec.~\ref{sec:eigenvalues}.
Thus, below the critical point the spectrum shows central and outer doublets associated with the photonic and atomic
branch  eigenvalues, respectively. The peaks of the photonic branch doublet merge as $\lambda\rightarrow\lambda_{\rm c}$,
forming a single narrow peak at $\nu=0$; within the linearized treatment the intensity under this peak diverges at
$\lambda=\lambda_{\rm c}$. Above the critical point a pair of doublets appears again. Far above the critical point
the photonic branch peaks approach detunings determined by the cavity mode resonance frequency, $\nu\simeq
\pm\omega=\pm1$, and linewidths (FWHM) $2\kappa=0.4$. The atomic branch peaks move linearly apart, following the
increasing Rabi frequency in the presence of the increasing mean intracavity field; they also become increasingly sharp.

Note that the symmetry of the spectra is ensured by energy conservation and the fact that, due to the symmetrical
nature of the atom-cavity coupling, photon emissions from the cavity can be associated with transitions to either
lower or higher internal energy states of the atom-cavity system.\\\\

\subsubsection{Probe transmission spectrum}

One may also examine the system by driving the cavity mode with a (weak) laser field and measuring the intensity of
the coherent transmission as a function of laser frequency. Such a measurement provides a rather direct 
probe of the energy level structure of the atom-cavity system; only when the laser frequency matches a system
resonance would substantial transmission be expected.

Analytically, we treat the measurement by adding a driving term, ${\cal E}_{\rm p}e^{-i\nu_{\rm p}t}$, to the equations
of motion for $\hat{a}$ and $\hat{c}$, where ${\cal E}_{\rm p}$ and $\nu_{\rm p}$ are the probe field amplitude and
frequency. Solving the equations of motion in frequency space as before, the coherent amplitude in transmission follows
straightforwardly from the coefficient of $\delta(\nu-\nu_{\rm p})$ in the solution for $\langle\tilde{a}_{\rm out}
(\nu)\rangle$. The transmitted probe intensity is thus found to be
\begin{widetext}
\begin{eqnarray}
T^{(1,2)}(\nu_{\rm p})=\kappa^2\left|\frac{\left[\kappa-i(\nu_{\rm p}+\omega_0)\right]\left[\nu_{\rm p}^2-\omega_0^2
/\left(\tilde{\mu}^{(1,2)}\right)^2\right]-2i\omega_0\lambda^2\tilde{\mu}^{(1,2)}}{\left[\kappa-i(\nu_{\rm p}-\omega_0)
\right]\left[\kappa-i(\nu_{\rm p} +\omega_0)\right]\left[\nu_{\rm p}^2-\omega_0^2/\left(\tilde{\mu}^{(1,2)}\right)^2
\right]+4\omega_0^2\lambda^2\tilde{\mu}^{(1,2)}}\right|^2, 
\end{eqnarray}
\end{widetext}
with $\tilde{\mu}^{(1)}$ and $\tilde{\mu}^{(2)}$ defined by Eq.~(\ref{eq:mu12}); the normalization is such that
the spectrum is a Lorentzian of width $2\kappa$ and unit height when $\lambda$ is set to zero.

\begin{figure}[h]
\begin{center}
\includegraphics[scale=0.4]{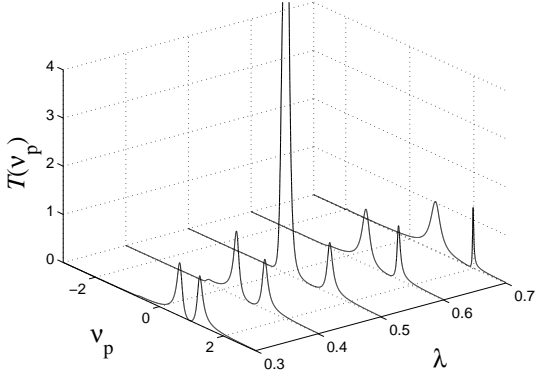}
\caption{Probe transmission spectrum $T(\nu_{\rm p})$ for various values of the coupling strength $\lambda$; for
$\omega=\omega_0=1$ and $\kappa=0.2$ ($\lambda_{\rm c}\simeq0.51$).}
\label{fig:probetransmission}
\end{center}
\end{figure}

A series of probe transmission spectra are plotted in Fig.~\ref{fig:probetransmission}, where we choose values of
coupling strength to correspond to Fig.~\ref{fig:fluorescence}. The spectra contain two principal peaks, one 
associated with the photonic and one with the atomic branch. Their behavior as a function of $\lambda$ replicates
the behavior displayed by the fluorescence.

\subsubsection{Homodyne spectra}

Homodyne spectra measure the fluctuation variances in frequency space of the output field quadrature amplitudes.
Quadrature operators are defined in time and frequency space, respectively, as
\begin{subequations}
\begin{eqnarray}
\hat{X}_{{\rm out},\theta}&=&\frac{1}{2}\left(\hat{a}_{\rm out}e^{-i\theta}+\hat{a}_{\rm out}^\dagger e^{i\theta}
\right),\\
\tilde{X}_{{\rm out},\theta}(\nu)&=&\frac{1}{2}\left[\tilde{a}_{\rm out}(\nu)e^{-i\theta}+\tilde{a}_{\rm out}^\dagger
(-\nu)e^{i\theta}\right],
\end{eqnarray}
where $\theta$ is the quadrature phase. The (normally-ordered) homodyne spectrum, $S_{{\rm out},\theta}(\nu)$,
is defined by the variance \cite{Collett84,Walls94}
\end{subequations}
\begin{eqnarray}
\langle\mkern2mu:\mkern-2mu\tilde{X}_{{\rm out},\theta}(\nu),\tilde{X}_{{\rm out},\theta}(\nu^\prime)\mkern-2mu
:\mkern2mu\rangle=S_{{\rm out},\theta}(\nu)\,\delta(\nu+\nu^\prime),
\end{eqnarray}
which we compute from the input-output relation (\ref{eq:io_relation}) and solutions (\ref{eq:a_nu}) and (\ref{eq:c_nu})
for the intracavity fields. Note that with the choice of normal ordering the vacuum noise level corresponds to
$S_{{\rm out},\theta}(\nu)=0$, while perfect quantum noise reduction corresponds to $S_{{\rm out},\theta}(\nu)=-1/4$.

\begin{figure}[t]
\begin{center}
\includegraphics[scale=0.4]{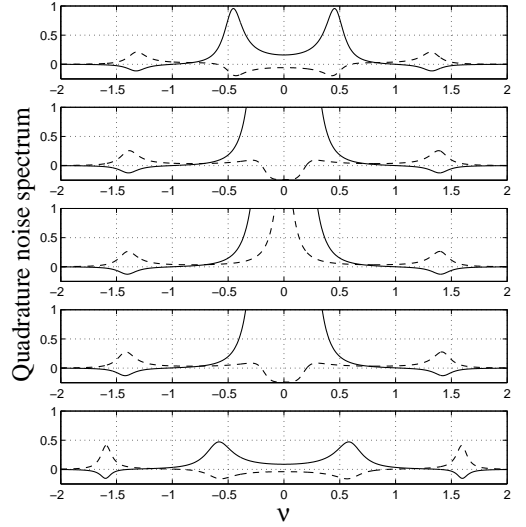}\\
\caption{Quadrature noise spectra $S_{{\rm out},\theta}(\nu)$ with $\theta=0$ (solid) and $\theta=\pi/2$ (dashed);
for $\omega=\omega_0=1$, $\kappa=0.2$ ($\lambda_{\rm c}\simeq0.51$), 
and $\lambda=0.4,0.49,0.505,0.52,0.6$ (top to bottom).}
\label{fig:squeezing}
\end{center}
\end{figure}

\begin{figure}[b]
\begin{center}
\includegraphics[scale=0.4]{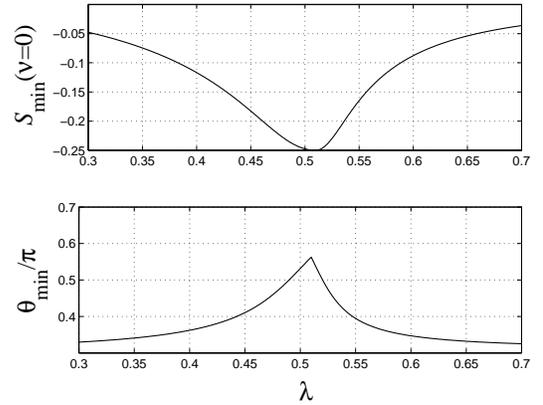}\\
\caption{Optimal squeezing at $\nu=0$ as a function of $\lambda$ (top) and the quadrature phase angle at which
the optimum occurs (bottom); for $\omega=\omega_0=1$ and $\kappa=0.2$ ($\lambda_{\rm c}\simeq0.51$).}
\label{fig:optimumsqueezing}
\end{center}
\end{figure}

Numerical results for $\theta=0$ and $\theta=\pi/2$ are presented in Fig.~\ref{fig:squeezing}. As the coupling strength
approaches $\lambda_{\rm c}$, the phase transition is signaled by a divergence of the quadrature amplitude flutuations
at $\nu=0$, similar to the behavior of the cavity fluorescence (Fig.~\ref{fig:fluorescence}). Nonetheless, there is an
optimal $\theta$ at each $\lambda$, for which near-perfect noise reduction occurs in the $(\theta+\pi/2)$-quadrature
at $\nu=0$. Figure~\ref{fig:optimumsqueezing} plots the optimal phase and  corresponding minimum quadrature variance
across the threshold region. As $\lambda\rightarrow\lambda_{\rm c}$, the optimal phase approaches $\theta_{\rm min}=
\tan^{-1}(\kappa/\omega)+\pi/2$.

Well above the critical point the noise level returns to the vacuum noise level at all frequencies, except close to
the atomic branch resonances at $\nu\simeq\pm\omega_0/\tilde{\mu}$. Here significant squeezing below the vacuum noise
level is found for the $\theta=0$ quadrature amplitude, with corresponding amplification of the fluctuations at
$\theta=\pi/2$. In fact, substantial squeezing of the atomic branch resonances occurs also for $\lambda<\lambda_{\rm c}$,
as seen from Fig.~\ref{fig:squeezing}. Although the bandwidth of this squeezing becomes increasingly narrow as $\lambda$
increases, the noise reduction on resonance actually approaches 100\%.

\subsubsection{Output field squeezing and atom-field entanglement}
\label{sec:squeeze}

For the parameter regime we have considered, the spectra presented exhibit distinct features that can be identified with
either the ``photonic'' or ``atomic'' branches. The lower frequency peaks are associated with the photonic branch and the
higher frequency peaks with the atomic branch. The corresponding photonic and atomic modes are formalized by a diagonalization
of Hamiltonians (\ref{eq:H1}) and (\ref{eq:H2}) via Bogoliubov transformations, as shown in \cite{Emary03b} and outlined in
Appendix A. If these modes are well separated in frequency, and $\kappa$ is sufficiently small, then one can also associate
with them what are essentially independent and uncorrelated output fields, $\hat{a}_{\rm out}^{\rm at}(t)$ and
$\hat{a}_{\rm out}^{\rm ph}(t)$; hence, we can relate their quadrature variances to the quadrature variances of linear
combinations of the ``bare'' internal atomic and cavity modes. In particular, in the normal phase, we find that the EPR
variance of Eq.~(\ref{eq:insep}) (with $\phi=\theta$) is approximately given by (Appendix A)
\begin{widetext}
\begin{equation}
\frac{2}{\kappa}\left(\langle\mkern3mu:\mkern-2mu\hat{X}_{{\rm out},\theta}^{\rm at},\hat{X}_{{\rm out},\theta}^{\rm at}
\mkern-2mu:\mkern3mu\rangle+\langle\mkern3mu:\mkern-2mu\hat{X}_{{\rm out},\theta+\pi/2}^{\rm ph},\hat{X}_{{\rm out},
\theta+\pi/2}^{\rm ph}\mkern-2mu:\mkern3mu\rangle \right)+1,
\label{eq:EPRvar_out}
\end{equation}
where the output field quadrature variances are calculated from integrals of the (normally-ordered) homodyne spectrum
over appropriate frequency ranges, i.e.,
\begin{subequations}
\begin{eqnarray}
\langle\mkern3mu:\mkern-2mu\hat{X}_{{\rm out},\theta+\pi/2}^{\rm ph},\hat{X}_{{\rm out},\theta+\pi/2}^{\rm ph}\mkern-2mu
:\mkern3mu\rangle &=& \frac{1}{2\pi}\int_{\{\nu_{\rm ph}\}}S_{{\rm out},\theta+\pi/2}(\nu)\,{\rm d}\nu,\\
\langle\mkern3mu:\mkern-2mu\hat{X}_{{\rm out},\theta}^{\rm at},\hat{X}_{{\rm out},\theta}^{\rm at}\mkern-2mu:\mkern3mu
\rangle &=& \frac{1}{2\pi}\int_{\{\nu_{\rm at}\}}S_{{\rm out},\theta}(\nu)\,{\rm d}\nu.
\end{eqnarray}
If one then considers the homodyne spectra plotted for $\lambda=0.4$ and $0.49$ in Fig.~\ref{fig:squeezing}, qualitatively,
these expressions allow entanglement to be inferred from the fact that $S_{{\rm out},0}(\nu)$ exhibits squeezing---i.e., is
negative---on the atomic branch while $S_{{\rm out},\pi/2}(\nu)$ exhibits squeezing on the photonic branch. Given the 
well-defined peaks and dips in the homodyne spectra  around $\theta=0,\,\pi/2$, we estimate (\ref{eq:EPRvar_out})
by
\end{subequations}
\begin{equation}
V_{\rm est}=\frac{2}{\kappa}\frac{1}{2\pi}\left\{\int_{S_{{\rm out},\theta}(\nu)<0} S_{{\rm out},\theta}(\nu)\,{\rm d}
\nu+\int_{S_{{\rm out},\theta+\pi/2}(\nu)<0}S_{{\rm out},\theta+\pi/2}(\nu)\,{\rm d}\nu \right\}+1.
\label{eq:Vest}
\end{equation}
\end{widetext}

\begin{figure}[h]
\begin{center}
\includegraphics[scale=0.4]{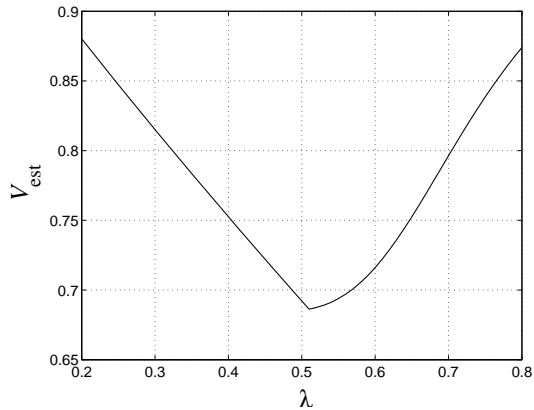}\\
\caption{Estimate of the EPR variance, $V_{\rm est}$, as a function of coupling strength; for $\omega=\omega_0=1$,
$\kappa=0.2$, and quadarture phase angle $\theta=\tan^{-1}(\kappa/\omega)\simeq0.2$.  The choice of $\theta$
minimizes $V_{\rm est}$ close to the threshold at $\lambda_{\rm c}\simeq0.51$.}
\label{fig:Vest}
\end{center}
\end{figure}

This quantity is plotted as a function of $\lambda$ in Fig.~\ref{fig:Vest}. For $\lambda<\lambda_{\rm c}$, it shows
rather good agreement with the EPR variance plotted in Fig.~\ref{fig:EPRvariance}; the agreement improves for decreasing
values of the decay rate $\kappa$. Above threshold, on the other hand, $V_{\rm est}$ can only be regarded as a good measure
of entanglement when $\lambda$ is quite close to $\lambda_{\rm c}$. In the superradiant phase, the relationships between
output field and internal mode operators are more complicated [compare Eqs.~(\ref{eq:phbelow})--(\ref{eq:atvarbelow}) and
(\ref{eq:phabove})--(\ref{eq:atvarabove}], but entanglement measures based on output field quadrature variances can still
be derived. The measures depend explicitly on $\lambda$, however, and cannot be directly related to the EPR variance of
Eq.~(\ref{eq:insep}), as was possible for the normal phase. Nevertheless, they do display a drop-off in the degree of
entanglement with increasing $\lambda$, consistent with that shown in Fig.~\ref{fig:EPRvariance}.

\section{Conclusion}

In this paper we have proposed a scheme for the realization of an elementary atom-light interaction Hamiltonian --
the so-called Dicke Model -- which should enable the observation and detailed study of a quantum phase transition
involving a collective atomic pseudo-spin and a single quantized mode of the electromagnetic field. While the optical
cavity-QED system considered is necessarily dissipative, due to cavity loss, the dissipation is a positive feature
providing a window through which one can monitor the system using standard quantum-optical measurement techniques.
As we have demonstrated, fluorescence, probe transmission, and squeezing spectra all provide detailed information
on the varying energy level structure of the Dicke Hamiltonian  and exhibit striking behavior in the vicinity of
the critical point.

We have focussed exclusively on the thermodynamic limit, with the number of atoms taken to infinity, where
fluctuations may be treated using a bosonic approximation for the collective atomic spin and linearization about
the semiclassical steady state. Finite-size systems are a natural consideration, both theoretically and
experimentally, and are of interest for examining scaling properties and deviations from the linearized model.
Indeed, in a regime of strong-coupling cavity QED (see, for example, \cite{McKeever03,Sauer04,Maunz05}) it could
be possible to realize the critical regime of the Dicke Model with just a few atoms. In such a case, issues of
quantum measurement (e.g., measurement backaction) arise, providing a further interesting avenue of investigation.

Finite-size systems and a full treatment of the Dicke model without linearization are also of importance for
studying the role of quantum entanglement in the vicinity of the phase transition. In the present paper we touched
only briefly on this subject, demonstrating that variance-based measures of atom-field entanglement can in principle
be determined from homodyne spectra of the cavity output field, thus enabling entanglement to be ``monitored''.
The proposed system clearly offers further exciting prospects for the study of entanglement in a quantum critical
system. For example, with additional light fields (possibly including other cavity modes) one could envisage making
independent measurements on the atomic ensemble to complement those made on the cavity field, enabling the explicit
determination of correlations and entanglement measures such as the EPR variance. Separately addressable atomic
sub-ensembles coupled to the same quantized cavity mode would also allow the measurement of entanglement between
different ``blocks'' of spins \cite{Latorre05}.

\begin{acknowledgments}
This work was supported by the Marsden Fund of the Royal Society of New Zealand. A.S.P. gratefully acknowledges support from the Institute for Quantum Information at the California Institute of Technology
and thanks the Quantum Optics Group of H.~J. Kimble for its hospitality.
\end{acknowledgments}

\vbox{\vskip1cm}
\appendix

\section{Normal modes and entanglement criteria}

\subsection{Normal phase}

The normal-phase Hamiltonian (\ref{eq:H1}) can be diagonalised in the form
(omitting constant terms)
\begin{eqnarray}
\hat{H}^{(1)}=\omega_{\rm ph}^{(1)}(\lambda)\hat{A}^\dagger\hat{A}+\omega_{\rm at}^{(1)}(\lambda)\hat{B}^\dagger\hat{B},
\end{eqnarray}
where $\omega_{\rm ph}^{(1)}(\lambda)$ and $\omega_{\rm at}^{(1)}(\lambda)$ are the normal mode frequencies, with
respective normal mode operators \cite{Emary03b}
\begin{widetext}
\begin{subequations}
\begin{eqnarray}
\hat{A}&=&\frac12\left(2\omega_0\omega_{\rm ph}^{(1)}\right)^{-1/2}\left[(\omega_{\rm ph}^{(1)}-\omega_0)
(\hat{a}^\dagger-\hat{b}^\dagger )+(\omega_{\rm ph}^{(1)}+\omega_0)(\hat{a}-\hat{b})\right],\label{eq:Amode}\\
\hat{B}&=&\frac12\left(2\omega_0\omega_{\rm at}^{(1)}\right)^{-1/2}\left[(\omega_{\rm at}^{(1)}-\omega_0)
(\hat{a}^\dagger+\hat{b}^\dagger)+(\omega_{\rm at}^{(1)}+\omega_0)(\hat{a}+\hat{b})\right],\label{eq:Bmode}
\end{eqnarray}
where $\omega=\omega_0$ has been assumed. The inverse relationship for the cavity mode operator $\hat{a}$ is
\end{subequations}
\begin{equation}
\hat{a}=\frac{1}{2}\left\{\left(2\omega_0\omega_{\rm ph}^{(1)}\right)^{-1/2}\left[(\omega_0-\omega_{\rm ph}^{(1)})
\hat{A}^\dagger+( \omega_0 + \omega_{\rm ph}^{(1)})\hat{A}\right]+\left(2\omega_0\omega_{\rm at}^{(1)}\right)^{-1/2}
\left[(\omega_0-\omega_{\rm at}^{(1)})\hat{B}^\dagger+(\omega_0+\omega_{\rm at}^{(1)})\hat{B}\right]\right\}.
\end{equation}
\end{widetext}
If the normal modes are well separated in frequency with linewidths much smaller than their separation, within the bandwidth
of the photonic mode the cavity mode contribution to the output field [Eq.~(\ref{eq:io_relation})] may be written as
\begin{equation}
\hat{a}\simeq\frac12\left(2\omega_0\omega_{\rm ph}^{(1)}\right)^{-1/2}\left[(\omega_0-\omega_{\rm ph}^{(1)})
\hat{A}^\dagger+(\omega_0+\omega_{\rm ph}^{(1)})\hat{A}\right],
\end{equation}
and within the bandwidth of the atomic mode as
\begin{equation}
\hat{a}\simeq\frac12\left(2\omega_0\omega_{\rm at}^{(1)}\right)^{-1/2}\left[( \omega_0-\omega_{\rm at}^{(1)})
\hat{B}^\dagger+(\omega_0+\omega_{\rm at}^{(1)})\hat{B}\right].
\end{equation}
Using these approximations, the input-output relation (\ref{eq:io_relation}), and Eqs.~(\ref{eq:Amode}) and (\ref{eq:Bmode}), 
one may derive approximate expressions for the output field quadrature operators in the specified frequency regions in
terms of ``bare'' cavity and atomic mode operators:
\begin{subequations}
\begin{eqnarray}
\hat{X}_{{\rm out},\theta}^{\rm ph}&\simeq&\sqrt{2\kappa}\mkern3mu(\hat{X}_a^\theta-\hat{X}_b^\theta)/2
-\hat{X}_{{\rm in},\theta}^{\rm ph},\label{eq:phbelow}\\
\hat{X}_{{\rm out},\theta}^{\rm at} &\simeq& \sqrt{2\kappa}\mkern3mu(\hat{X}_a^\theta+\hat{X}_b^\theta )/2
-\hat{X}_{{\rm in},\theta}^{\rm at}.
\end{eqnarray}
It follows that, in the normal phase, the normally-ordered output field variances can be directly related to the
internal mode EPR variances \cite{Collett84}:
\end{subequations}
\begin{subequations}
\begin{eqnarray}
\langle\mkern2mu:\!\hat{X}_{{\rm out},\theta}^{\rm ph},\hat{X}_{{\rm out},\theta}^{\rm ph}\!:\mkern2mu\rangle
&\simeq&\frac\kappa2\langle\mkern2mu :\!\hat{X}_a^\theta - \hat{X}_b^\theta ,\hat{X}_a^\theta - \hat{X}_b^\theta
\!:\mkern2mu\rangle,\\
\langle\mkern2mu:\!\hat{X}_{{\rm out},\theta}^{\rm at},\hat{X}_{{\rm out},\theta}^{\rm at}\!:\mkern2mu\rangle
&\simeq&\frac\kappa2\langle\mkern2mu:\!\hat{X}_a^\theta+\hat{X}_b^\theta,\hat{X}_a^\theta+\hat{X}_b^\theta
\!:\mkern2mu\rangle,\mkern30mu
\label{eq:atvarbelow}
\end{eqnarray}
where a vacuum field input has been assumed. Then, adopting the entanglement criterion from \cite{Duan00}, entanglement
between the cavity and atomic modes can be inferred whenever the inequality
\end{subequations}
\begin{equation}
\langle\mkern3mu:\mkern-2mu\hat{X}_{{\rm out},\theta}^{\rm at},\hat{X}_{{\rm out},\theta}^{\rm at}\mkern-2mu
:\mkern3mu\rangle+\langle\mkern3mu:\mkern-2mu\hat{X}_{{\rm out},\theta+\pi/2}^{\rm ph},\hat{X}_{{\rm out},
\theta+\pi/2}^{\rm ph}\mkern-2mu:\mkern3mu\rangle<0
\label{eq:output_inequ}
\end{equation}
is satisfied.

\subsection{Superradiant phase}
The superradiant-phase Hamiltonian (\ref{eq:H2}) can be diagonalised in similar fashion in the form
(omitting constant terms)
\begin{eqnarray}
\hat{H}^{(2)}=\omega_{\rm ph}^{(2)}(\lambda)\hat{C}^\dagger\hat{C}+\omega_{\rm at}^{(2)}(\lambda)\hat{D}^\dagger\hat{D},
\end{eqnarray}
where $\omega_{\rm ph}^{(2)}$ and $\omega_{\rm at}^{(2)}$ are the above threshold normal mode frequencies (for
linearization around either of the above threshold steady states), and the respective normal mode operators
are given by the somewhat more complicated expressions \cite{Emary03b}
\begin{widetext}
\begin{subequations}
\begin{eqnarray}
\hat{C}&=&\frac{1}{2}\left\{\frac{\cos\gamma^{(2)}}{\sqrt{\omega_0\omega_{\rm ph}^{(2)}}}\left[(\omega_{\rm ph}^{(2)}
-\omega_0)\hat{c}^\dagger+(\omega_{\rm ph}^{(2)}+\omega_0) \hat{c}\right]-\frac{\sin\gamma^{(2)}}{\sqrt{\tilde{\omega}_0
\omega_{\rm ph}^{(2)}}}\left[(\omega_{\rm ph}^{(2)}-\tilde{\omega}_0)\hat{d}^\dagger+(\omega_{\rm ph}^{(2)}
+\tilde{\omega}_0)\hat{d} \right]\right\},\\
\hat{D}&=&\frac{1}{2} \left\{\frac{\sin\gamma^{(2)}}{\sqrt{\omega_0\omega_{\rm at}^{(2)}}}\left[(\omega_{\rm at}^{(2)}
-\omega_0) \hat{c}^\dagger+(\omega_{\rm at}^{(2)}+\omega_0) \hat{c}\right]+\frac{\cos\gamma^{(2)}}
{\sqrt{\tilde{\omega}_0\omega_{\rm at}^{(2)}}}\left[(\omega_{\rm at}^{(2)}-\tilde{\omega}_0)\hat{d}^\dagger
+(\omega_{\rm at}^{(2)}+\tilde{\omega}_0)\hat{d}\right]\right\},
\end{eqnarray}
\end{subequations}
\end{widetext}
with
\begin{equation}
\tan(2\gamma^{(2)})=2\tilde{\mu}^2(1-\tilde{\mu}^2)^{-1},
\end{equation}
\begin{equation}
\tilde{\omega}_0=\omega_0(1+\tilde{\mu}^{-1})/2,
\end{equation}
where, once again, the resonance condition $\omega=\omega_0$ has been assumed. These expressions do not allow for
as simple a relationship between output field and internal mode quadrature variances to be written down. Nevertheless, 
following the same arguments as before, one can write
\begin{subequations}
\begin{eqnarray}
\hat{X}_{{\rm out},\theta}^{\rm ph}&\simeq&\sqrt{2\kappa}\mkern3mu\hat X_{cd}^\theta-\hat{X}_{{\rm in},\theta}^{\rm ph},
\label{eq:phabove}\\
\hat{X}_{{\rm out},\theta}^{\rm at}&\simeq&\sqrt{2\kappa}\mkern3mu\hat Y_{cd}^\theta-\hat{X}_{{\rm in},\theta}^{\rm at} ,
\end{eqnarray}
where
\end{subequations}
\begin{widetext}
\begin{subequations}
\begin{eqnarray}
\hat X_{cd}^\theta&=&\cos^2(\gamma^{(2)})\hat{X}_c^\theta-\cos(\gamma^{(2)})\sin(\gamma^{(2)})\mkern-2mu
\left[\cos\theta\sqrt{\frac{\omega_0}{\tilde{\omega}_0}}\mkern3mu\hat{X}_d^{\theta=0}+\sin\theta
\sqrt{\frac{\tilde{\omega}_0}{\omega_0}}\mkern3mu\hat{X}_d^{\theta=\pi/2}\right],\\
\hat Y_{cd}^\theta&=&\sin^2(\gamma^{(2)})\hat{X}_c^\theta+\cos(\gamma^{(2)})\sin(\gamma^{(2)})\mkern-2mu
\left[\cos\theta\sqrt{\frac{\omega_0}{\tilde{\omega}_0}}\mkern3mu\hat{X}_d^{\theta=0}+\sin\theta 
\sqrt{\frac{\tilde{\omega}_0}{\omega_0}}\mkern3mu \hat{X}_d^{\theta=\pi/2}\right].
\end{eqnarray}
For these more complicated linear superpositions of the internal mode operators it is still possible to derive inseparability
criteria based on their variances. In particular, following \cite{Giovannetti03} one can show that a sufficient condition for
the inseparability of the system state is given by
\end{subequations}
\begin{equation}
V_1 \equiv 
\frac{\left\langle(\Delta\hat X_{cd}^{\theta+\pi/2})^2\right\rangle+\left\langle(\Delta\hat Y_{cd}^{\theta})^2\right\rangle}
{\cos^2(\gamma^{(2)})\sin^2(\gamma^{(2)})}<1,
\end{equation}
or in stronger form
\begin{equation}
V_2 \equiv
\frac{\left\langle(\Delta\hat X_{cd}^{\theta+\pi/2} )^2\right\rangle\left\langle(\Delta\hat Y_{cd}^{\theta} )^2\right\rangle}
{\frac{1}{4}\cos^4(\gamma^{(2)})\sin^4(\gamma^{(2)})}<1.
\end{equation}
The required variances can be deduced from the (normally-ordered) photonic and atomic output field quadrature variances by
inverting the relations
\begin{subequations}
\begin{equation}
\langle\mkern3mu:\hat{X}_{{\rm out},\theta+\pi/2}^{\rm ph},\hat{X}_{{\rm out},\theta+\pi/2}^{\rm ph}:\mkern3mu\rangle
\simeq2\kappa\left\{\mkern-2mu\left\langle(\Delta\hat X_{cd}^{\theta+\pi/2} )^2\right\rangle-\frac{1}{4}\left[\cos^4
(\gamma^{(2)})+\cos^2(\gamma^{(2)})\sin^2(\gamma^{(2)})\mkern-2mu\left(\frac{\omega_0}{\tilde{\omega}_0}\sin^2\theta
+\frac{\tilde{\omega}_0}{\omega_0}\cos^2\theta \right)\right]\mkern-2mu\right\},
\end{equation}
and
\begin{equation}
\langle\mkern3mu:\hat{X}_{{\rm out},\theta}^{\rm at},\hat{X}_{{\rm out},\theta}^{\rm at}:\mkern3mu\rangle \simeq 2\kappa
\left\{\mkern-2mu\left\langle(\Delta\hat Y_{cd}^{\theta} )^2 \right\rangle-\frac{1}{4}\left[\sin^4(\gamma^{(2)})+\cos^2
(\gamma^{(2)}) \sin^2(\gamma^{(2)})\mkern-2mu\left(\frac{\omega_0}{\tilde{\omega}_0}\cos^2\theta+\frac{\tilde{\omega}_0}
{\omega_0}\sin^2\theta \right)\right]\mkern-2mu\right\}.\label{eq:atvarabove}
\end{equation}
Numerical examples of $V_1$ and $V_2$ versus $\lambda$ are shown in Fig.~\ref{fig:V12}. The output field quadrature variances used
were computed via numerical integration of the homodyne spectra from Section IV.E.4. The computed $V_1$ and $V_2$ display a decay
in the degree of entanglement with increasing $\lambda$, consistent with the internal mode EPR variance of Fig.~\ref{fig:EPRvariance}.
\end{subequations}
\begin{figure}[h]
\begin{center}
\includegraphics[scale=0.45]{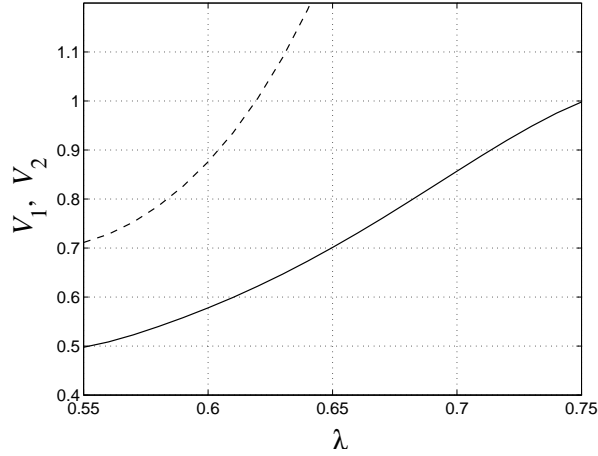}\\
\caption{Entanglement measures $V_1$ (dashed) and $V_2$ (solid) versus coupling strength; for $\omega=\omega_0=1$, $\kappa=0.2$,
and $\theta=0$.}
\label{fig:V12}
\end{center}
\end{figure}
\end{widetext}

\end{document}